\documentclass{optica-article}

\journal{opticajournal} 

\articletype{Research Article}

\usepackage{lineno}
\usepackage{bm}
\usepackage{amsmath}

\begin{document}

\title{Choice of optical transformation for photonic circuit wavefront sensors}

\author{Jonathan Lin,\authormark{1}}

\address{\authormark{1}NASA Ames Research Center, 1 Moffett Field, CA 94035, USA}

\email{\authormark{*}jonathan.w.lin@nasa.gov} 


\begin{abstract*} 
Coronagraph designs which use photonic integrated circuits have the highest theoretical throughput for off-axis signals, and therefore the highest potential exoplanet yield for future high-contrast direct imaging campaigns. Using the rejected starlight, the photonic integrated circuit may also provide simultaneous wavefront sensing, allowing for the correction of non-common-path aberrations. This work considers how a photonic circuit should be configured to maximize its sensitivity to phase aberrations.
Two cases are considered: in the first, the photonic circuit is coupled directly to an electric field in a piecewise manner, while in the second, the circuit is coupled to the field via an optical mode sorter. 
In either case, this work constructs a unitary matrix which can be applied by a photonic circuit to produce maximum sensitivity.

\end{abstract*}


\section{Introduction}
Photonic integrated circuits (PICs) are a type of active waveguide which can, in principle, apply any unitary transformation to an incoming complex-valued electric field. Combined with their compact footprint and inherent optomechanical stability \cite{roadmap}, PICs present one pathway to achieving the stringent $10^{10}$ suppression of starlight required to directly image an exoEarth in the visible wavelengths \cite{traub}. In fact, previous works \cite{rus,sirbu1,sirbu2} have shown that PIC-based coronagraphy can reach the fundamental limit in exoplanetary light throughput \cite{guyoncoro}, improving exoplanet yields. 

Because coronagraphic contrast is strongly impacted by the flatness of the incoming wavefront, and therefore the quasistatic instrumental aberrations that come with all real instruments, future coronagraphs will also require active wavefront control. The expectation for a PIC-based coronagraph is that at least part of this wavefront control should be supplied by the circuit itself, working in tandem with a conventional adaptive optics system and using the starlight rejected by the coronagraphic stage of the PIC: such a setup could sense and correct non-common-path aberrations \cite{ncpa}, one of the main challenges in deepening coronagraphic contrast.

PIC wavefront sensors (WFSs) should also be able to perform wavefront sensing at the ``fundamental limit'' -- pending definition of this limit -- in the same way that a PIC-based coronagraph can preserve the maximum amount of off-axis exoplanet light while simultaneously blocking the on-axis star. Considering that coronagraphic WFSs will likely need to correct for small amounts of residual wavefront error at high precision, this work considers the limit in the sensitivity to phase aberrations; that said, limits on metrics such as the dynamic range, e.g. those caused by WFS nonlinearity \cite{me24}, may also be useful, especially in ground-based astronomy where wavefront errors are larger. Beyond coronagraphy, highly sensitive PICs may also be of interest in interferometric imaging (e.g. VLTI/GRAVITY \cite{benisty} and GLINT \cite{glint}), particularly for fringe tracking, as well as in free-space optical communications and remote sensing.

Under the scalar approximation, define the phase of the electric field across the telescope aperture as $\bm{\phi}$, which is expanded as $\bm{\phi}=\sum_k a_k \bm{\phi}_k$ over a basis $\bm{\phi}_k$ covering the aperture and normalized to unit RMS amplitude (e.g. the Zernike modes for a circular pupil). Also denote the intensity output of a generic WFS as the function $\bm{I}(\bm{a}) \propto |U e^{i\boldsymbol{\phi}}|^2$, where $U$ is the unitary operation applied by the WFS and the components of $\bm{I}$ are the intensity outputs (i.e. pixel values) of the WFS, assumed to be corrupted solely by photon noise. The goal of this work is to find the matrix $U$ which yields the smallest uncertainty on the estimate of the mode amplitudes, denoted $\hat{a}_k$. This can be framed as a maximization of sensitivity. Similar to previous works on the Zernike WFS \cite{chamb1,chamb2,haffert}, define the single-channel sensitivity in WFS output $j$ to an aberration mode $\boldsymbol{\phi}_k$ as
\begin{equation}
    S_j(U,\boldsymbol{\phi}_k) = \Bigg|\dfrac{\partial I_j}{\partial a_k}\Big|_{\bm{a}=0}  \Bigg|/\sqrt{I_j}
\end{equation}
and the total sensitivity for mode $k$ as the single-channel sensitivities added in quadrature:
\begin{equation}\label{eq:sens}
    \mathcal{S}(U,\boldsymbol{\phi}_k) = \sqrt{\sum_j S_j^2(U,\boldsymbol{\phi}_k)}.
\end{equation}
While the above is sometimes derived from the Fisher information \cite{haffert,Trzaska:24}, it can also be derived in the context of estimation theory. From this perspective, the sensitivity is the inverse of the uncertainty in $\hat{\bm{a}}$ due to photon noise, where $\hat{\bm{a}}$ is the maximum-likelihood estimate of the true mode amplitudes $\bm{a}$ (see Appendix \ref{ap} for a more complete treatment). The maximum sensitivity for any unitary optical system, as defined in Eq. \ref{eq:sens}, is 2. For a more rigorous proof, see \cite{pat}; however, the following gives a heuristic reasoning (c.f. \cite{chamb3}). Suppose there are two single-moded optical beams, represented by the complex numbers (phasors) $a$ and $b$. The total power is normalized to 1: $|a|^2+|b|^2=1$. The interference of the two beams gives an intensity pattern $I = |a+b|^2 = |a|^2 + |b|^2 + 2 \textrm{Re}(a^*b)$, where $*$ denotes complex conjugate. The sensitivity, for instance with respect to $b$, is
\begin{equation}\label{eq:lim}
    \mathcal{S} = \Big| \dfrac{\partial I}{\partial b}\Big| = 2\Big| \dfrac{\partial }{\partial b} \textrm{Re}(a^*b)\Big|  \leq 2,
\end{equation}
where equality is reached for $|a|\rightarrow 1$ and $|b|\rightarrow 0$ (and both completely real or imaginary); note that this maintains the normalization condition. The best sensitivity is achieved when the beam intensities are maximally unbalanced: this is typical of optical homodyne detection \cite{OH}, where sensitivity improves with the intensity of the local oscillator. Incidentally, this is also why the Zernike WFS gains sensitivity at sufficiently high spatial frequency when the phase dimple size is increased, as observed in \cite{chamb1}: the intensity response of the Zernike WFS is a two-wave interference pattern, and the larger phase dimple produces a brighter reference wave.

The goal of this paper is to derive the $N\times N$ unitary transformations that achieve the maximum sensitivity. This is somewhat opposite to the coronagraphic application, which tries to minimize the sensitivity with respect to low order phase aberrations \cite{guyoncoro,rus} while maintaining sufficient off-axis planet throughput. Such transformations could be applied by a PIC WFS -- assumed to be separate from but on the same chip as the PIC coronagraph, to avoid the aforementioned tension -- via Reck decomposition \cite{reck} or \cite{clements}. In comparison with other existing coronagraphic WFSs, a PIC WFS would give true common-path sensing, and could be more sensitive (e.g. compared to the Zernike WFS, especially at low spatial frequency).

This work considers two scenarios: in the first, the PIC WFS is directly coupled to a spatially discretized electric field, while in the second, light from the telescope is first processed by an optical mode sorter. These configurations represent the two extremes in how a telescope might be coupled to a PIC.

\begin{figure}
    \centering
    \includegraphics[width=\linewidth]{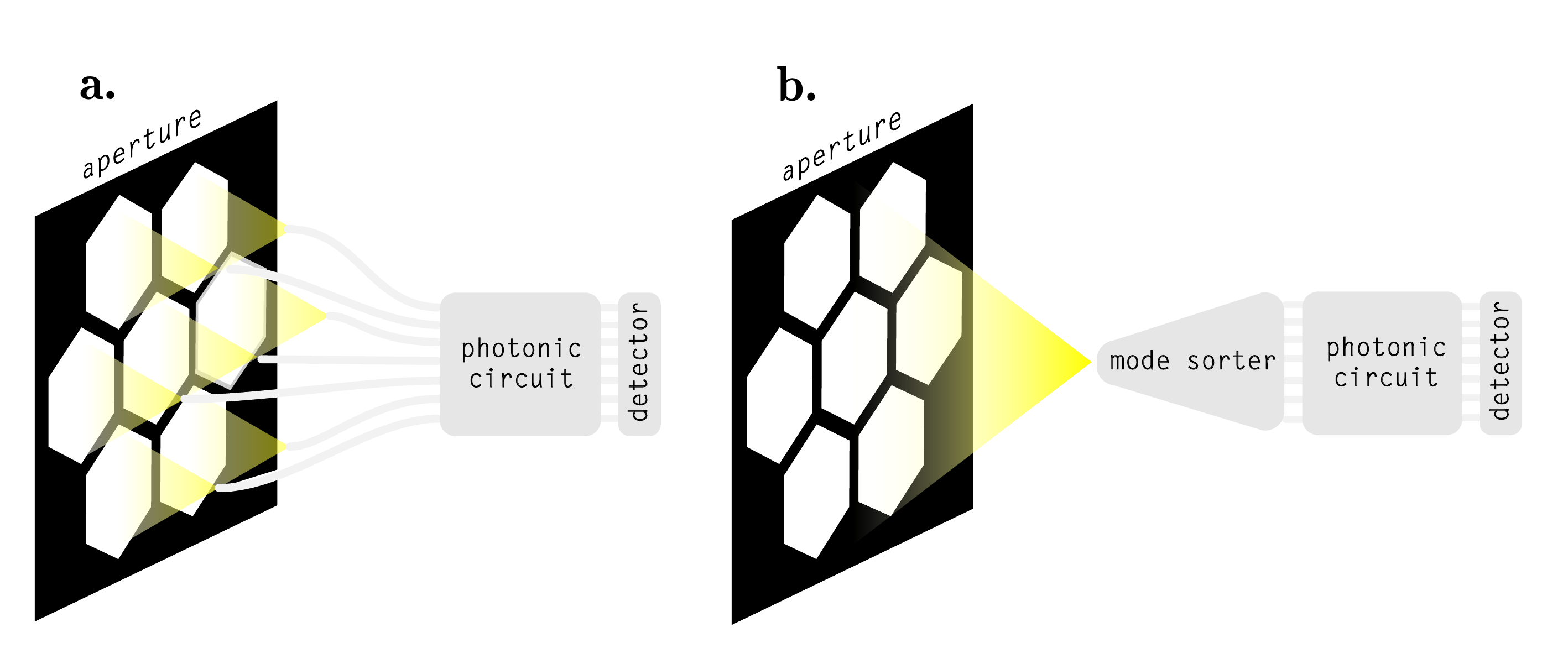}
    \caption{\textbf{a.} Coupling of the PIC with a spatially discretized wavefront. This coupling could be accomplished in many ways, including via optical fibers (as drawn), or a microlens array which focuses light into an on-chip array of couplers, and could occur in a pupil plane (as drawn) or any other plane. Note that exoplanet detection and characterization with a PIC does not strictly require image formation. For fragmented apertures, each fragment could also be divided into multiple subapertures/channels. \textbf{b.} Diagram of the mode-sorting setup; the mode sorter splits the electric field into orthogonal modes and couples each into a PIC channel. This mode sorter could operate in the pupil plane or focal plane (as drawn). In either case, the aperture may be more general than the one drawn here.}
    \label{fig:diagram}
\end{figure}

\section{Coupling to a spatially discretized field}\label{sec:disc}
As shown in Figure \ref{fig:diagram}a, a PIC couples light from the telescope in a spatially piecewise manner. For mathematical convenience, this work assumes that coupling occurs in the pupil plane, so that the telescope aperture is split into $N$ subapertures of equal size, each of which is coupled into a single channel of the circuit without loss. In this scenario, the PIC acts like an $N$-beam interferometer which measures the relative local piston displacements between subapertures. In principle, coupling could occur in any optical plane, the only difference being that the additional free-space propagation to this plane will rotate the phase aberration basis, making it complex-valued. To gain intuition, first consider $N=2$, which corresponds to a two-beam interferometer.
\subsection{A toy model: $N=2$}\label{ssec:N2}
Consider a unit-power input wavefront represented by the complex-valued vector $s_\textrm{in}$, where
\begin{equation}
    \bm{s}_\textrm{in} = \dfrac{1}{\sqrt{2}}\begin{bmatrix}
        e^{i \phi_1}\\
        e^{i\phi_2}
    \end{bmatrix}.
\end{equation}
The two beams are mixed by some $2\times 2$ unitary matrix $U$; then, the output intensity is
\begin{equation}
    \bm{I} = |\bm{s}_\textrm{out}|^2=|U\bm{s}_\textrm{in}|^2
\end{equation}
where $|\bm{s}|^2 = \bm{s}^* \odot \bm{s}$; $\odot$ denotes elementwise (Hadamard) product. As an illustrative example, consider choosing $U$ as the $2\times 2$ Hadamard matrix:
\begin{equation}
    U_{\textrm{Had},2} \equiv \dfrac{1}{\sqrt{2}}
    \begin{bmatrix}
        1 & 1 \\
        1 & -1 
    \end{bmatrix}.
\end{equation}
The resulting intensity is
\begin{equation}\label{eq:had}
    \bm{I}_{\textrm{Had},2} = \dfrac{1}{2}\begin{bmatrix}
        1 + \cos(\Delta \phi) \\
        1 - \cos(\Delta \phi)
    \end{bmatrix}
\end{equation}
which is completely symmetric about $\Delta \phi \equiv \phi_1-\phi_2$: this setup cannot tell us in which direction to phase the subapertures, and is not a good choice as a sensor. A better choice is
\begin{equation}\label{eq:lou}
    U_{2} = \dfrac{1}{\sqrt{2}}
    \begin{bmatrix}
        1 & i \\
        i & 1 
    \end{bmatrix},
\end{equation}
which yields the intensity output
\begin{equation}\label{eq:loudon}
    \bm{I}_{2} = \dfrac{1}{2}\begin{bmatrix}
        1 + \sin(\Delta \phi) \\
        1 - \sin(\Delta \phi)
    \end{bmatrix}.
\end{equation}
This result can also be obtained from the above by subtracting $\pi/2$ from the argument of the cosines in Eq. \ref{eq:had}, which is the same as changing the reference phase.

Application of Eq. \ref{eq:sens} to $U_2$, taking $a=\Delta\phi$, yields a phase sensitivity of 1, consistent with other references on two-beam interferometry \cite{su2}. In fact, a direct application of Eq. \ref{eq:sens} to Eq. \ref{eq:loudon} shows that the phase sensitivity is 1 for all $\phi \neq \pi/2$ ($\pm$ multiples of $\pi$), at which points the sensitivity is undefined, corresponding to a Hadamard-like configuration. However, the choice of aberration basis implicit in taking $a=\Delta\phi$ (essentially a Cartesian basis) is not orthogonal to the global\footnote{``Global'' piston is used to identify the piston mode over the entire telescope aperture.} piston mode, and would result in an apparent loss of sensitivity. The next subsection derives one construction of a maximally sensitive WFS matrix, for a better choice of phase aberration basis.

\begin{figure}
    \centering
    \includegraphics[width=\linewidth]{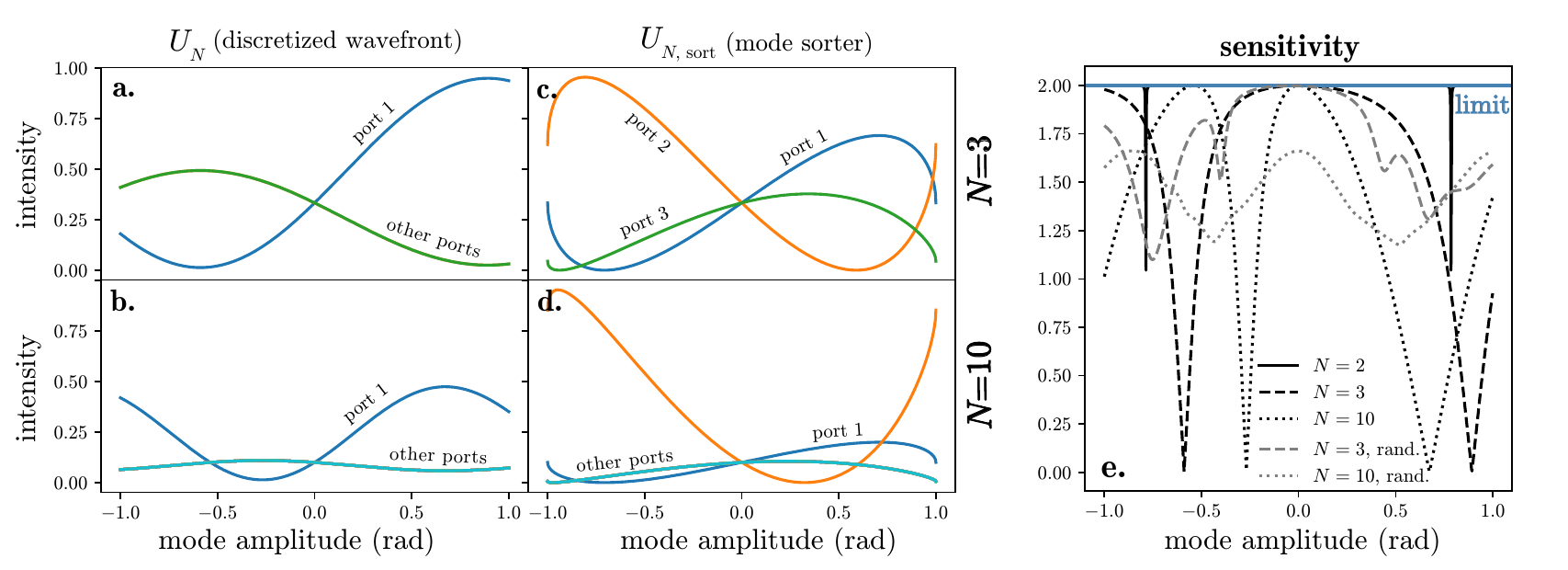}
    \caption{\textbf{a, b}: Intensity response of $U_{3}$ and $U_{10}$ to the $\boldsymbol{\phi}_{\mathrm{SP},k}$ mode basis (the local subaperture piston modes, orthogonalized against the global piston). \textbf{c, d}: Intensity response of $U_{3,\textrm{sort}}$ and $U_{10,\textrm{sort}}$ when scanning over an aberration mode. While scanning, the total power of the electric field is kept at 1. Note that the behavior near the edges is not meaningful because it violates the small angle approximation; however, the behavior near 0 phase is accurate. \textbf{e}: Numerically computed sensitivity of $U_{N}$ as a function of reference mode amplitude. Spikes in the $N=2$ curve correspond to point discontinuities. For comparison, each gray curve corresponds to a matrix which obtained the highest sensitivity out of $10^4$ random samples. The blue curve is the theoretical limit,which $U_{N,\textrm{sort}}$ also reaches at $\bm{\phi}=0$.}
    \label{fig:int}
\end{figure}

\subsection{A maximally sensitive WFS matrix for $N\geq2$}
One strategy to develop a good unitary transformation for $N\geq2$ is to try to ``extend'' the $U_2$ matrix derived earlier. Consider the eigendecomposition of $U_2$: 
\begin{equation}
    U_2 = U_{\textrm{Had},2} \,\textrm{diag}(1,-i)\,U_{\textrm{Had},2}^T
\end{equation}
where diag\{$\lambda_1,\lambda_2,\cdots$\} is a diagonal matrix with the given entries; a global phase factor is omitted for clarity. The $U_2$ operation sorts the incoming light into the ``cophased'' electric field mode,\footnote{Electric field modes are complex-valued and should have unit norm, while phase aberration modes are real-valued and should have unit RMS amplitude. Under the small angle approximation, a phase aberration mode basis, combined with a normalization factor, may be taken as a basis for the electric field.} which is $[1,1]^T/\sqrt{2}$, and the orthonormal ``aberrated'' mode $[1,-1]^T/\sqrt{2}$, applies a $\pi/2$ phase shift to the aberrated mode, and then changes back to the original basis (c.f. \cite{chamb3,Trzaska:24}). This strategy is extended as follows:
\begin{enumerate}
    \item Change into a basis composed of a completely cophased electric field mode, $[1,\cdots]^T/\sqrt{N}$, and orthonormal modes in the kernel of this cophased mode.
    \item Apply a $\pi/2$ phase shift (all positive or negative) to all modes except the cophased mode.
    \item Change back to the original basis.
\end{enumerate}
One choice of electric field basis is the discrete Fourier mode basis, which exist for arbitrary $N$ and contains a global piston mode. These modes are the columns of the discrete Fourier transform matrix $F$:
\begin{equation}
    F_{jk} = \dfrac{1}{\sqrt{N}} e^{2 \pi i (j-1)(k-1)/N}.
\end{equation}
Then, one choice of unitary transformation for arbitrary $N$ is
\begin{equation}\label{eq:UN}
    U_{N} \equiv F \, \textrm{diag}\{1,i,i,\cdots\} \,F^\dagger
\end{equation}
which turns out to be a unitary circulant matrix, where the columns and rows are cyclic permutations of each other; $\dagger$ denotes conjugate transpose. From here, the sensitivity can be computed under suitable choice of a phase aberration basis. Assume a basis of phase modes $\boldsymbol{\phi}_k$ normalized to unit RMS amplitude, where $\boldsymbol{\phi}_1=[1,\dots]^T$ corresponds to the global piston mode and all $\boldsymbol{\phi}_{k>1}$ are orthogonal to $\boldsymbol{\phi}_1$. Under the linear approximation, the incoming wavefront may be expanded as 
\begin{equation}
    \bm{s}_{\rm in} = \dfrac{1}{\sqrt{N}} e^{i\boldsymbol{\phi}} \approx \frac{1}{\sqrt{N}}\left( \boldsymbol{\phi}_1 + i \sum_k a_k \boldsymbol{\phi}_k\right).
\end{equation}
The intensity output of a wavefront sensor performing some operation $U$, to first order in the mode amplitudes, can be written as 
\begin{equation}
    \bm{I} = |U\bm{s}_{\rm in}|^2 \approx \dfrac{1}{N}|U\boldsymbol{\phi}_1|^2 - \dfrac{2}{N} \mathrm{Im}\left[ (U \boldsymbol{\phi}_1)^* \odot \left(\sum_k a_k U \boldsymbol{\phi}_k\right) \right].
\end{equation}
Differentiating the above gives
\begin{equation}
    \dfrac{\partial I_j}{\partial a_k} = -\dfrac{2}{N}\mathrm{Im}\left[ (U\boldsymbol{\phi}_1)_j^*\,(U\boldsymbol{\phi}_k)_j \right].
\end{equation}
This gives the following form for the sensitivity with respect to mode $k$:
\begin{equation}\label{eq:sens2}
\begin{split}
    \mathcal{S}(U,\boldsymbol{\phi}_k)  &= \sqrt{\sum_j \left(\dfrac{\partial I_j}{\partial a_k}\right)^2/I_j} \\
    &= 2\sqrt{\dfrac{1}{N}\sum_j \left(\mathrm{Im}\left[ (U\boldsymbol{\phi}_1)_j^*\,(U\boldsymbol{\phi}_k)_j \right]\right)^2}\\
\end{split}
\end{equation}
Eq. \ref{eq:sens2} shows that a WFS operation $U$ cannot be maximally sensitive to phase and amplitude aberrations simultaneously, since under the small angle approximation, phase and amplitude aberrations are rotated 90$^\circ$ from each other in the complex plane (c.f. \cite{Trzaska:25}). Next, substitute in $U_N$. 
By construction, the operation $U_N$ applies a relative $\pi/2$ phase shift between the global piston mode and all other modes. Therefore
\begin{equation}
\begin{split}
     \mathcal{S}(U_N,\boldsymbol{\phi}_k) &= 2\sqrt{\dfrac{1}{N}\sum_j \left(\mathrm{Im}\left[ (U_N\boldsymbol{\phi}_k)_j \right]\right)^2} \\
     &= \begin{cases}
         0 &  k=1\\
         2 & k>1
     \end{cases}. 
\end{split}
\end{equation}
The last equality follows from the assumption that the phase aberration modes have unit RMS amplitude. Thus, a PIC applying the $U_N$ transformation achieves the maximum sensitivity to a basis of $N-1$ phase aberration modes (one ``slot'' is taken by global piston), so long as the phase aberration modes are chosen to be orthogonal to global piston and normalized to unit RMS amplitude.
Figure \ref{fig:int}a and b show the intensity response of $U_N$ for $N=3$ and $N=10$, respectively, while panel e show the sensitivity as a function of the reference mode amplitude for different values of $N$. As $N$ increases, the range over which high sensitivity is achieved becomes narrower: the WFS becomes more nonlinear as the light in the global piston mode is progressively split. As a sanity check, panel e also shows the sensitivities of matrices produced by random optimization, where the matrix with the highest sensitivity was selected from $10^4$ random samples. Interestingly, while these matrices do not achieve the highest peak sensitivity, they remain sensitive over a larger range than the $U_N$ matrices, suggesting future opportunities for joint optimization.

The sensitivity in each WFS channel to a ``local'' subaperture piston mode, corresponding to a phase shift in one of the subapertures, can also be computed. Orthogonalizing against global piston and normalizing to unit RMS amplitude, these modes, denoted $\boldsymbol{\phi}_{\mathrm{SP},k}$, can be written as
\begin{equation}
    \boldsymbol{\phi}_{\mathrm{SP},k} = 
    \begin{bmatrix}
        \sqrt{N-1} \\
        -1/\sqrt{N-1} \\
        \vdots
    \end{bmatrix}
\end{equation}
where the subaperture $k$ has been reordered to the top for clarity. Note that these modes do not form an orthogonal basis and therefore have non-zero crosstalk. The sensitivity in channel $j$ for this mode basis is 
\begin{equation}
S_j(U_N,\boldsymbol{\phi}_{\mathrm{SP},k}) =
\begin{cases}
    2\sqrt{\dfrac{N-1}{N}} & j =k \\
    \dfrac{2}{\sqrt{N(N-1)}} & j\neq k
\end{cases}
\end{equation}
which sums in quadrature to 2, as expected.

\section{Coupling to an optical mode sorter}
This section considers a PIC WFS which accepts light from an optical mode sorter that couples the aberrated wavefront into different optical modes and passes the light from each into a channel of the circuit. This setup is shown in Figure \ref{fig:diagram}b. Such a mode sorter could be realized with photonic lanterns \cite{birks}, holographic phase masks \cite{Neil:00,holo}, multiplane light conversion \cite{mplc}, or vortex phase masks \cite{vortex}, and in principle could operate in any optical plane; given the result of the previous discussion, the mode sorter should at the very least isolate the global piston mode, or whatever corresponds to global piston in the optical plane of the sorter. Mathematically, many of the results in this section will look similar to the previous section, which is unsurprising: considering the decomposition of $U_N$ in Eq. \ref{eq:UN}, the $F^\dagger$ term acts as a mode sorter. However, there is a practical difference between the two approaches. If the dominant aberration modes of an optical system are known, then an optical mode sorter designed for those modes could reduce the number of channels required on the PIC.

\subsection{Choice of basis modes}
Without loss of generality, suppose the mode sorter is located in the pupil plane and has a projection matrix $P^\dagger$. Then, the electric field after the mode sorter may be written as
\begin{equation}
    \bm{s}_{\rm sort}=P^\dagger \bm{s}_{\rm in} = P^\dagger\left(e^{i\boldsymbol{\phi}}/\sqrt{A}\right) \approx \dfrac{1}{\sqrt{A}}\left( P^\dagger \boldsymbol{\phi}_1 +  i\sum_k a_k P^\dagger\boldsymbol{\phi}_k \right)
\end{equation}
where $A$ is the area of the aperture, $a_k$ are real-valued coefficients, and $\bm{\phi}_k$ is now an orthogonal basis of phase aberration modes, again assumed to be normalized to unit RMS amplitude with $k=1$ corresponding to the global piston mode. Taking the columns of $P$ as $\boldsymbol{\phi}_k/\sqrt{A}$ gives
\begin{equation}
    \bm{s}_{\rm sort} = \bm{a}_0
    + i \bm{a} + o(\bm{a}^2)
\end{equation}
with $\bm{a}_0=[1,0,0,\cdots]^T$. 
If the mode sorter is not in the pupil plane, then the mode basis selected by the columns of $P$ should be chosen as the electric field modes corresponding to each phase aberration mode, propagated to whatever plane the sorter is in; $a_k$ may still be taken to be purely real with suitable phasing of the basis.
The corresponding intensity post-WFS, assuming the WFS applies a unitary operation $U$, is
\begin{equation}
    \bm{I} = \Big| U \bm{s}_{\rm sort}\Big|^2 =\Big| U \left(\bm{a}_0+i\bm{a}+\cdots\right)\Big|^2
\end{equation}
or expanding in indexed notation:
\begin{equation}\label{eq:int}
    I_j= I_{0,j} + 2 \textrm{Im}\left(U_{j1}  
    \sum_k {U}_{jk}^*a_k\right) + o(\bm{a}^2).
\end{equation}
where the reference intensity $\bm{I}_0$ is defined as 
\begin{equation}
    \bm{I}_0 \equiv |U\bm{a}_0|^2.
\end{equation}
The derivatives of the intensity $\bm{I}$ with respect to aberration mode amplitudes $a_k$ are
\begin{equation}\label{eq:focalslope}
    \dfrac{\partial I_j}{\partial a_k} = 2 \,\textrm{Im}\left(U_{j1}{U}_{jk}^*\right) + o(\bm{a}^2).
\end{equation}

\subsection{Matrix construction} \label{ssec:Nfocal}
A general strategy to construct a matrix $U$ which maximizes the derivatives in Eq. \ref{eq:focalslope}, and hence the sensitivity, is the following:
\begin{enumerate}
    \item The first column and row of $U$ are set to $1/\sqrt{N}$. This ensures that the PIC evenly splits light when there is no phase aberration, while obeying unitarity.
    \item The remaining elements form an $(N-1)\times (N-1)$ circulant matrix, where the largest element is on the diagonal and all other elements have the same modulus. To determine the values, note that the sum of all remaining columns of $U$ must be 0, to maintain orthogonality with the first column, and the norm of all remaining columns must be 1.
    \item All columns after the first are multiplied through by $i$, giving the $\pi/2$ phase shift.
\end{enumerate}
In practice, the columns of $U$ after the first may be computed numerically, e.g. by applying the \texttt{scipy} function \texttt{linalg.nullspace} to the global piston mode $[1,1,\cdots]^T/\sqrt{N}$, and multiplying the result through by $i$. Denote the matrix constructed in this way as $U_{N,\textrm{sort}}$.
Some algebra shows that the largest single-channel sensitivity is 
\begin{equation}\label{eq:singsensfocal}
    S_j (U_{N,\textrm{sort}},\boldsymbol{\phi}_k)= 2\dfrac{(N-2)\sqrt{N}-1}{\sqrt{N}(N-1)}\quad \textrm{for}\quad k>1.
\end{equation}
The sensitivity in the first channel is 
\begin{equation}
    S_{1}(U_{N,\textrm{sort}},\boldsymbol{\phi}_k) = \dfrac{2}{\sqrt{N}} \quad \textrm{for}\quad k>1
\end{equation}
and all other channels have sensitivity
\begin{equation}
    S_{j}(U_{N,\textrm{sort}},\boldsymbol{\phi}_k) =  2\dfrac{(N-2)\sqrt{N}-N+2}{\sqrt{N}(N-2)(N-1)}\quad \textrm{for} \quad j>1\,;\,j\neq k.
\end{equation}
Using the above, it can be shown that the total sensitivity for all modes with $k>1$ is 2:
\begin{equation}
    \mathcal{S}(U_{N,\textrm{sort}},\boldsymbol{\phi}_k) = 2 \quad \textrm{for}\quad k>1.
\end{equation}
In the limit $N\rightarrow \infty$, this sensitivity is obtained in a single channel, as seen in the asymptotic behavior of Eq. \ref{eq:singsensfocal}. Figures \ref{fig:int}c and d show the intensity response of the above matrix for $N=3$ and 10, respectively, while scanning over the amplitude of an aberration mode. 
\section{Discussion}
This work recovers the fact that a WFS which is optimally sensitive to phase aberrations must sort out the global piston mode, apply a relative $\pi/2$ phase shift, and reinterfere the modes. This sorting behavior may be performed by the circuit itself, in the case that the circuit channels are coupled in a piecewise manner to a spatially discretized wavefront, or with a dedicated mode sorter; a WFS operation that reaches maximum sensitivity is provided in both cases. While the two approaches are mathematically similar, a mode sorter may be practically preferable if it can be designed to sort the dominant aberration modes expected in the physical system; otherwise, the spatial sampling of the wavefront by the PIC WFS must be sufficiently high so that the physical aberration modes are well-resolved. Considering the challenges in PIC fabrication and light efficiency, the approaches that yield the simpler designs may be preferable. Such a system could provide true common-path wavefront sensing for a PIC coronagraph with maximal sensitivity to any phase aberration mode, something that is not possible with existing coronagraphic WFSs.

Looking forwards, there are many areas for improvement in PIC wavefront sensing beyond sensitivity optimization. First, also active wavefront correction can also be implemented within the circuit, instead of relying completely on the slower deformable mirror \cite{diab}; this can be considered a part of ``dark hole digging'' for photonic coronagraphy, as demonstrated in \cite{carson}. Second, for ground-based astronomy and other applications dealing with larger amounts of wavefront error, dynamic range may be more important than sensitivity. Optimization of a PIC with respect to dynamic range is much more open-ended and challenging than the sensitivity optimization performed here \cite{me24}. Third, it may be beneficial to measure the complex electric field at the circuit output, instead of intensity, similar to how VLTI/GRAVITY uses ABCD beam combiners to measure complex interferometric visibility\cite{benisty}. A similar scheme for a PIC WFS could improve linearity, since it essentially removes the squaring step inherent in an intensity measurement, but would also require a doubling of the circuit's outputs.\footnote{Interestingly, the total sensitivity of the $4\times 2$ ABCD matrix from \cite{benisty}, as computed with Eq. \ref{eq:sens}, is the same as the $U_2$ matrix developed in \S\ref{ssec:N2} (ignoring the point discontinuities). That said, the contexts for interferometric imaging and phase aberration sensing are not quite the same.} Lastly, while this work derived a unitary matrix that obtains maximum sensitivity for a given basis of phase-only aberration modes, nothing bars us from changing the unitary matrix dynamically. For quasistatic aberrations it may be advantageous to concentrate flux into only a few channels of the circuit and sense aberration modes one at a time, thereby driving down photon noise, instead of trying to measure all phase aberrations in one shot. Other approaches which dynamically modulate the WFS matrix could also give better phase constraints in a manner analogous to phase diversity, and optimally sense both phase and amplitude aberrations.

\section{Conclusion}
This work derives an $N\times N$ unitary matrix which can be applied by an $N$-channel photonic circuit in order to obtain optimal phase aberration sensitivity, under two configurations: piecewise coupling of the circuit inputs to a discretized wavefront, and coupling via an optical mode sorter which separates out the global piston aberration. In both cases, the sensitivity phase aberrations reaches the limit of 2. Therefore, a PIC WFS could in principle provide common-path sensing for a PIC coronagraph with maximal sensitivity across a designated phase aberration basis.

\appendix
\section{Sensitivity in the context of estimation theory}\label{ap}

The goal of this section is show how WFS sensitivity naturally arises in an estimation theory framework. Denote the response matrix $M_{jk} \equiv (\partial I_j/\partial a_k)|_{\bm{a}=0}$ and $\bm{I}_0$ as the reference intensity when no aberrations or noise are present. To first order, the measurements produced by the WFS are 
\begin{equation}
    \bm{I} = \bm{I}_0 + M \bm{a} + \bm{n}
\end{equation}
where $\bm{n}$ is a vector of random variables representing measurement noise. In the case of uniform Gaussian noise, which is a good approximation for photon noise if all output channels of the WFS are bright and have similar intensity, the maximum likelihood estimate of $\bm{a}$ is given by
\begin{equation}
    \hat{\bm{a}} = M^+ \left(\bm{I} - \bm{I}_0\right),
\end{equation}
where $M^+$ is the pseudoinverse of $M$. The error propagation formula gives the following covariance matrix for $\hat{\bm{a}}$:  
\begin{equation}
\Sigma_{\hat{\bm{a}}} = M^+ \Sigma_{\bm{I}} M^{+ T}.
\end{equation}
Assuming photon noise, $\Sigma_{\bm{I}} = \textrm{diag}(\bm{I})$. Applying the properties of the pseudoinverse gives
\begin{equation}
\begin{split}
\Sigma_{\hat{\bm{a}}} &= (\Sigma_{\hat{\bm{a}}}^+)^+\\
&= \left(M^T \dfrac{1}{\textrm{diag}(\bm{I})} M\right)^+.
\end{split}
\end{equation}
Presuming that the covariance is diagonal, this gives 
\begin{equation}
\sigma_{\hat{a}_k}^2  = \dfrac{1}{\sum_j\left(\dfrac{\partial I_j}{\partial a_k}\right)^2/I_j} = \dfrac{1}{\mathcal{S}^2(\boldsymbol{\phi}_k)}.
\end{equation}
Therefore, the uncertainty in the mode amplitude estimates due to photon noise is the inverse of the sensitivity. The covariance will be diagonal if $M^+ M^{+T}$ is diagonal and the photon noise is uniform. The former can be enforced with appropriate choice of mode basis, and the latter is consistent with all the WFS operations derived in this work, since the operations were derived to evenly split light when no phase aberrations are present. The latter assumption may also be relaxed by using a weighted least-squares estimator for $\bm{a}$.

\begin{backmatter}

\bmsection{Acknowledgment}
The author would like to thank Ruslan Belikov and Michael Fitzgerald for helpful discussions on sensitivity calculations and photonic integrated circuits. This research was partly supported by cooperative agreement 80NSSC24M0197 between the NASA Ames Research Center and the SETI Institute. 
\\\\
This work was carried out at NASA Ames Research Center. Any opinions, findings, and conclusions or
recommendations expressed in this article are those of the authors and do not necessarily reflect the views of the National Aeronautics and Space Administration.

\bmsection{Disclosures}
The authors declare no conflicts of interest.

\bmsection{Data availability} No data were generated or analyzed in the presented research.

\end{backmatter}


\bibliography{refs}

\begin{thebibliography}{10}
\newcommand{\enquote}[1]{``#1''}

\bibitem{roadmap}
N.~Jovanovic, P.~Gatkine, N.~Anugu, \emph{et~al.}, \enquote{2023 astrophotonics roadmap: pathways to realizing multi-functional integrated astrophotonic instruments,} {\protect\JournalTitle{Journal of Physics: Photonics}} \textbf{5}, 042501 (2023).

\bibitem{traub}
W.~A. {Traub} and B.~R. {Oppenheimer}, \enquote{{Direct Imaging of Exoplanets},} in \emph{Exoplanets,}  S.~{Seager}, ed. (University of Arizona Press, 2010), pp. 111--156.

\bibitem{rus}
R.~Belikov, D.~Sirbu, J.~B. Jewell, \emph{et~al.}, \enquote{{Theoretical performance limits for coronagraphs on obstructed and unobstructed apertures: how much can current designs be improved?}} in \emph{Techniques and Instrumentation for Detection of Exoplanets X,}  vol. 11823 S.~B. Shaklan and G.~J. Ruane, eds., International Society for Optics and Photonics (SPIE, 2021), p. 118230W.

\bibitem{sirbu1}
D.~Sirbu, R.~Belikov, K.~Fogarty, \emph{et~al.}, \enquote{{AstroPIC: near-infrared photonic integrated circuit coronagraph architecture for the Habitable Worlds Observatory},} in \emph{Space Telescopes and Instrumentation 2024: Optical, Infrared, and Millimeter Wave,}  vol. 13092 L.~E. Coyle, S.~Matsuura, and M.~D. Perrin, eds., International Society for Optics and Photonics (SPIE, 2024), p. 130921T.

\bibitem{sirbu2}
D.~Sirbu, R.~Belikov, E.~Bendek, \emph{et~al.}, \enquote{{AstroPIC II: overview of technology development for a near-infrared photonic integrated coronagraph for the Habitable Worlds Observatory},} in \emph{UV/Optical/IR Space Telescopes and Instruments: Innovative Technologies and Concepts XII,}  vol. 13623 J.~W. Arenberg and H.~P. Stahl, eds., International Society for Optics and Photonics (SPIE, 2025), p. 1362309.

\bibitem{guyoncoro}
O.~Guyon, E.~A. Pluzhnik, M.~J. Kuchner, \emph{et~al.}, \enquote{Theoretical limits on extrasolar terrestrial planet detection with coronagraphs,} {\protect\JournalTitle{The Astrophysical Journal Supplement Series}} \textbf{167}, 81 (2006).

\bibitem{ncpa}
{Martinez, P.}, {Loose, C.}, {Aller Carpentier, E.}, and {Kasper, M.}, \enquote{Speckle temporal stability in xao coronagraphic images,} {\protect\JournalTitle{A \& A}} \textbf{541}, A136 (2012).

\bibitem{me24}
J.~{Lin} and M.~P. {Fitzgerald}, \enquote{{Nonlinear techniques for few-mode wavefront sensors},} {\protect\JournalTitle{Applied Optics}} \textbf{63}, 8748 (2024).

\bibitem{benisty}
{Benisty, M.}, {Berger, J.-P.}, {Jocou, L.}, \emph{et~al.}, \enquote{An integrated optics beam combiner for the second generation vlti instruments,} {\protect\JournalTitle{A\&A}} \textbf{498}, 601--613 (2009).

\bibitem{glint}
B.~R.~M. Norris, N.~Cvetojevic, T.~Lagadec, \emph{et~al.}, \enquote{First on-sky demonstration of an integrated-photonic nulling interferometer: the glint instrument,} {\protect\JournalTitle{Monthly Notices of the Royal Astronomical Society}} \textbf{491}, 4180--4193 (2019).

\bibitem{chamb1}
{Chambouleyron, V.}, {Fauvarque, O.}, {Sauvage, J-F.}, \emph{et~al.}, \enquote{Variation on a zernike wavefront sensor theme: Optimal use of photons,} {\protect\JournalTitle{A\&A}} \textbf{650}, L8 (2021).

\bibitem{chamb2}
{Chambouleyron, V.}, {Fauvarque, O.}, {Plantet, C.}, \emph{et~al.}, \enquote{Modeling noise propagation in fourier-filtering wavefront sensing, fundamental limits, and quantitative comparison,} {\protect\JournalTitle{A\&A}} \textbf{670}, A153 (2023).

\bibitem{haffert}
S.~Haffert, J.~Males, and O.~Guyon, \enquote{{Reaching the fundamental sensitivity limit of wavefront sensing on arbitrary apertures with the Phase Induced Amplitude Apodized Zernike Wavefront Sensor (PIAA-ZWFS).}} in \emph{{Adaptive Optics for Extremely Large Telescopes 7th Edition},}  ({ONERA}, Avignon, France, 2023).

\bibitem{Trzaska:24}
J.~Trzaska and A.~Ashok, \enquote{Approaching the quantum limit of wavefront sensing with spatial mode sorting,} in \emph{Optica Imaging Congress 2024 (3D, AOMS, COSI, ISA, pcAOP),}  (Optica Publishing Group, 2024), p. CF1B.2.

\bibitem{pat}
C.~Paterson, \enquote{Towards practical wavefront sensing at the fundamental information limit,} {\protect\JournalTitle{Journal of Physics: Conference Series}} \textbf{139}, 012021 (2008).

\bibitem{chamb3}
V.~{Chambouleyron}, J.~K. {Wallace}, R.~{Jensen-Clem}, and B.~{Macintosh}, \enquote{{Coronagraph-based wavefront sensors for the high Strehl regime},} {\protect\JournalTitle{Optics Express}} \textbf{32}, 47706 (2024).

\bibitem{OH}
B.~L. Schumaker, \enquote{Noise in homodyne detection,} {\protect\JournalTitle{Opt. Lett.}} \textbf{9}, 189--191 (1984).

\bibitem{reck}
M.~Reck, A.~Zeilinger, H.~J. Bernstein, and P.~Bertani, \enquote{Experimental realization of any discrete unitary operator,} {\protect\JournalTitle{Phys. Rev. Lett.}} \textbf{73}, 58--61 (1994).

\bibitem{clements}
W.~R. Clements, P.~C. Humphreys, B.~J. Metcalf, \emph{et~al.}, \enquote{Optimal design for universal multiport interferometers,} {\protect\JournalTitle{Optica}} \textbf{3}, 1460--1465 (2016).

\bibitem{su2}
B.~Yurke, S.~L. McCall, and J.~R. Klauder, \enquote{Su(2) and su(1,1) interferometers,} {\protect\JournalTitle{Phys. Rev. A}} \textbf{33}, 4033--4054 (1986).

\bibitem{Trzaska:25}
J.~Trzaska and A.~Ashok, \enquote{Fundamental limits to phase and amplitude estimation in the high-strehl regime,}  (2025).

\bibitem{birks}
T.~A. {Birks}, I.~{Gris-S{\'a}nchez}, S.~{Yerolatsitis}, \emph{et~al.}, \enquote{{The photonic lantern},} {\protect\JournalTitle{Advances in Optics and Photonics}} \textbf{7}, 107 (2015).

\bibitem{Neil:00}
M.~A.~A. Neil, M.~J. Booth, and T.~Wilson, \enquote{New modal wave-front sensor: a theoretical analysis,} {\protect\JournalTitle{J. Opt. Soc. Am. A}} \textbf{17}, 1098--1107 (2000).

\bibitem{holo}
G.~P. Andersen, L.~C. Dussan, F.~Ghebremichael, and K.~Chen, \enquote{{Holographic wavefront sensor},} {\protect\JournalTitle{Optical Engineering}} \textbf{48}, 085801 (2009).

\bibitem{mplc}
G.~Labroille, B.~Denolle, P.~Jian, \emph{et~al.}, \enquote{Efficient and mode selective spatial mode multiplexer based on multi-plane light conversion,} {\protect\JournalTitle{Opt. Express}} \textbf{22}, 15599--15607 (2014).

\bibitem{vortex}
J.~{Trzaska} and A.~{Ashok}, \enquote{{Zernike Mode Sorting with Vortex Phase Filters: Perfect Coronagraphs and Ideal Wavefront Sensors},} {\protect\JournalTitle{arXiv e-prints}} arXiv:2510.21981 (2025).

\bibitem{diab}
M.~Diab, R.~Cheriton, J.~Taylor, \emph{et~al.}, \enquote{{Experimental demonstration of photonic phase correctors based on grating coupler arrays and thermo-optic shifters},} in \emph{Adaptive Optics Systems IX,}  vol. 13097 K.~J. Jackson, D.~Schmidt, and E.~Vernet, eds., International Society for Optics and Photonics (SPIE, 2024), p. 130972K.

\bibitem{carson}
C.~G. Valdez, Z.~Sun, A.~R. Kroo, \emph{et~al.}, \enquote{High-contrast nulling in photonic meshes through architectural redundancy,} {\protect\JournalTitle{Opt. Lett.}} \textbf{50}, 3660--3663 (2025).

\end{thebibliography}






\end{document}